\documentclass[aps,onecolumn,amsmath,showpacs,superscriptaddress,floatfix,nofootinbib,nopreprintnumbers]{revtex4-2}

\usepackage{verbatim}
\usepackage[T1]{fontenc}
\usepackage[utf8]{inputenc}
\usepackage[american]{babel}
\usepackage{epsfig}
\usepackage{graphicx}
\usepackage{booktabs}
\usepackage{multirow}
\usepackage{dcolumn}
\usepackage{amsmath}
\usepackage{amssymb}
\usepackage{mathtools}
\usepackage{amsfonts}
\usepackage{amssymb}
\usepackage{epstopdf}
\usepackage{bm}
\usepackage{siunitx}
\usepackage{braket}
\usepackage{enumitem}
\usepackage{soul}
\usepackage[table]{xcolor}
\usepackage{color}
\usepackage{transparent}
\usepackage{pifont}
\usepackage[normalem]{ulem}

\definecolor{navyblue}{rgb}{0.0, 0.0, 0.5}
\definecolor{royalblue}{rgb}{0.25, 0.41, 0.88}
\definecolor{cadmiumgreen}{rgb}{0.0, 0.42, 0.24}
\definecolor{blue-violet}{rgb}{0.54, 0.17, 0.89}
\definecolor{darkviolet}{rgb}{0.58, 0.0, 0.83}
\definecolor{orange(colorwheel)}{rgb}{1.0, 0.5, 0.0}

\usepackage[bookmarks=true,bookmarksnumbered=true,colorlinks=true,urlcolor=royalblue, citecolor=blue,linkcolor=royalblue,breaklinks=true]{hyperref}
\hypersetup{
    colorlinks=true, 
    linkcolor=royalblue, 
    citecolor=royalblue}

\usepackage{booktabs}
\usepackage{multirow}
\usepackage{dcolumn}
\usepackage{colortbl}

\definecolor{magenta(process)}{rgb}{1.0, 0.0, 0.56}

\definecolor{darkspringgreen}{rgb}{0.09, 0.45, 0.27}

\definecolor{royalblue(web)}{rgb}{0.25, 0.41, 0.88}

\begin{document}

\title{Redshift Space Distortions corner interacting Dark Energy}

\author{Pietro Ghedini}
\email{pietro.ghedini3@studio.unibo.it}
\affiliation{Dipartimento di Fisica e Astronomia, Università di Bologna, via Irnerio 46, 40126 Bologna, Italy}

\author{Rasmi Hajjar}
\email{rasmi.hajjar@ific.uv.es}

\author{Olga Mena}
\email{omena@ific.uv.es}

\affiliation{Instituto de F\'{i}sica Corpuscular (IFIC), University of Valencia-CSIC, Parc Cient\'{i}fic UV, C/ Cate\-dr\'{a}tico Jos\'{e} Beltr\'{a}n 2, E-46980 Paterna, Spain}

\date{\today}
\preprint{}

\begin{abstract}
Despite the fact that the $\Lambda$CDM model has been highly successful over the last few decades in providing an accurate fit to a broad range of cosmological and astrophysical observations, different intriguing tensions and anomalies emerged at various statistical levels. Given the fact that the dark energy and the dark matter sectors remain unexplored, 
the answer to some of the tensions may rely on modifications of these two dark sectors. 
This manuscript explores the important role of the growth of structure in constraining non-standard cosmologies. In particular, we focus on  the interacting dark energy (IDE) scenario, where dark matter and dark energy interact non-gravitationally. We aim to place constraints on the phenomenological parameters of these alternative models, by considering different datasets related to a number of cosmological measurements, to achieve a complementary analysis. A special emphasis is devoted to redshift space distortion measurements (RSD), whose role in constraining beyond the standard paradigm models has not been recently highlighted. These observations indeed have a strong constraining power, rendering all parameters to their $\Lambda$CDM canonical values, and therefore leaving little room for the IDE models explored here.
\end{abstract}


\maketitle

\section{Introduction}
\label{sec:intro}
Cosmological observations have provided us clear evidences for the existence  of both a dark energy and a dark matter components, but their nature and putative interactions, beyond the pure gravitational one, remain unknown. Since observations allow it, one could extend the $\Lambda$CDM model by introducing a new non-gravitational interaction in the dark sector, i.e. between dark energy and dark matter. That is, while the strength of interactions between ordinary matter and the dark energy fields is severely constrained by observations~\cite{Carroll_1998}, interactions among the dark sectors are still allowed. Therefore, over the last several years, the possibility of an  interaction between the dark matter and the dark energy fluids has been thoroughly investigated using different available cosmological  observations~\cite{Barrow_2006, V_liviita_2008, Gavela:2009cy, Gavela_2010, Li_2014, Salvatelli_2014, Wang_2014, Casas_2016, de_Bruck_2016, Yang_2016, Kumar_2016, van_de_Bruck_2017, Pan_2017, Yang_2017, Sharov_2017, Di_Valentino_2017, Kumar_2017, Mifsud_2017, van_de_Bruck_2018, Yang_2017a, Yang_2018, Pan_2018, Yang_2018a, Yang_2018b, Yang_2018c, Yang_2018d, Li:2018jiu, Yang_2018e, yang2019, Yang_2020, Li_2020, Di_Valentino_2020, Oikonomou_2019, Di_Valentino_2021, Martinelli_2019, Kumar_2019, Cheng_2020, Lucca_2020, Cid_2021, Sa_2020, BeltranJimenez_2021, Wang_2022, Kumar_2021, Nunes_2021, Di_Valentino_2020a, Yang_2019, Yang_2020a, Di_Valentino_2020b, G_mez_Valent_2020, Yang_2020b, Pan_2020, Yao_2020, yao2020, mifsud2019, Pan_2019, Paliathanasis_2019, Yang_2019a, Yang_2020c, Sa_2021, kang2021, da_Fonseca_2022, Yang_2021, lucca2021, Lucca_2021, Gariazzo_2022, Paliathanasis_2021, mawas2021, Potting_2022, Harko_2022, Nunes_2022, Ghodsi_2023, Thipaksorn_2022, Landim_2022, yao2024, Gomez_Valent_2022}, see also the reviews of Refs.~\cite{Bolotin_2015, Wang_2016}. This ``dark coupling'', based on models of coupled quintessence~\cite{wetterich1994, Amendola_2000, Amendola_2000a, MANGANO_2003}, could affect significantly the evolution history of the Universe and the evolution of the density perturbations. The presence of such a coupling could alleviate the \emph{``coincidence'' problem} since the introduction of a coupling could stabilise the ratio of the two dark components during the entire expansion history. Interacting dark energy-dark matter cosmologies~\cite{Amendola_2000a}  may not only provide the scenario where to alleviate the \emph{``why now?'' problem} but also could help in solving some of the existing cosmological tensions, such as the one between CMB estimates~\cite{planck18VI, Aiola_2020, Dutcher_2021} and SH0ES (Supernovae and $H_0$ for the Equation of State of dark energy) measurements of the Hubble constant~\cite{Riess_2022, Riess_2022a}, with a significance of $\sim 5.3\sigma$ (see also Refs.~\cite{Verde_2019, Knox_2020, Di_Valentino_2021a, Jedamzik_2021, Riess_2019, Di_Valentino_2021b, Di_Valentino_2021c, Perivolaropoulos_2022, Sch_neberg_2022, Shah_2021, DiValentino2022, Abdalla_2022, Krishnan_2021, Anchordoqui_2021, Colgin_2022, Colgain:2022rxy, Naidoo_2024}).  
 
The basic underlying idea in these theories relies on the possible non-gravitational interaction between dark matter and dark energy. Such an interaction can be characterized by a continuous flow of energy and/or momentum between these two dark sectors. This energy flow modifies the expansion history of the Universe both at the background and perturbation levels. The interaction function representing the continuous flow of energy and/or momentum between the dark matter and  dark energy sectors is also known as the coupling function, and it is the main feature of interacting dark energy theories: once the interaction function is prescribed, the dynamics of the Universe can be determined either analytically or numerically. Despite the fact that in the literature it is usually assumed a pure phenomenological approach for the dark sectors coupling, a class of interaction functions can be derived from the field theory perspective~\cite{de_Bruck_2015, B_hmer_2015, B_hmer_2015a, Gleyzes_2015, D_Amico_2016, Pan_2020a, chatzidakis2022}. Consequently some  interaction functions can also find a well-defined motivation, with the possibility of translating the bounds of the phenomenological parameters into bounds of the parameters of the proposed theory.    

Very recently, the DESI collaboration has presented new high-precision Baryon Acoustic Oscillation (BAO) measurements \cite{desiIII, desiIV} and new cosmological results \cite{desiVI}. The new DESI results point to a richer dark energy sector than that expected within the minimal $\Lambda$CDM scenario. Profiting from this exiting evidence, the recent work of Ref.~\cite{giarè2024} has shown a preference for interactions
exceeding the $95\%$ confidence level (CL). The very same study concludes that high
and low redshift observations can be equally or better
explained in interacting dark energy cosmologies than in the $\Lambda$CDM framework and also provide higher
values of $H_0$, compatible with SH0ES observations. DESI data may therefore imply non-standard cosmological scenarios and it is mandatory to fully explore these schemes in light of high and low redshift data. 

Interacting dark energy cosmologies are known to modify significantly the growth rate of structure. Measurements of the growth rate are at reach thank to Redshift Space Distortions (RSD), which provide another low-redshift observable to constrain non-canonical late time physics via the quantity $f\sigma_8$, being $f$ the growth rate of structure and $\sigma_8$ the matter clustering parameter. In the following, we analyze a class of interacting dark energy models in light of the most recent publicly available cosmological data. 

The structure of the paper is as follows. Section~\ref{sec-efe} contains an introduction to coupled dark matter-dark energy cosmologies. The cosmological data sets and the methodology are both described in Sec.~\ref{sec:method}. The bounds on the dark sector coupling arising from the different data sets considered here are presented in Sec.~\ref{sec:results}, and we present our conclusions and final remarks in Sec.~\ref{sec:conclusions}.

\section{Interacting dark sector scenarios}
\label{sec-efe}

We consider a homogeneous and isotropic description of our Universe, which is well described by the Friedmann-Lema\^{i}tre-Robertson-Walker (FLRW) metric

\begin{equation}
d{\rm s}^2 =  - dt^2 + a^2 (t) \left[\frac{dr^2}{1-k r^2} + r^2 \left(d\theta ^2 + \sin^2 \theta d \phi^2 \right) \right]~, 
\label{eq:flrw_metric}
\end{equation}
expressed in terms of the comoving coordinates $(t, r, \theta, \phi)$, where $a (t)$ refers to the scale factor of the Universe and $k$ is the curvature scalar. The curvature scalar may take three distinct values, $k =\{0,~+1,~-1\}$, to represent three different geometries of the Universe, spatially flat, closed and open, respectively.  In the following, we shall work in a flat Universe with an extra interaction between the pressure-less dark matter/cold dark matter and the dark energy components, governed by the conservation equation  
\begin{equation}
\nabla^\mu (T^{\rm dm}_{\mu\nu} + T^{\rm de}_{\mu\nu})= 0~,\label{eq:conservation}
\end{equation}
\noindent which can be decoupled into two equations with the introduction of an interaction function $Q (t)$ as follows
\begin{align}
\dot{\rho}_{\rm dm} + 3 H \rho_{\rm dm} &= Q (t)~, \label{eq:dm_background_evolution}\\
\dot{\rho}_{\rm de} + 3 H (1+w) \rho_{\rm de}  &= - Q (t)~;\label{eq:de_background_evolution}
\end{align}
\noindent where $H \equiv \dot{a}/a$ is the Hubble rate of a FLRW flat Universe, $\rho_{\rm dm (de)}$ are the time-dependent dark matter (dark energy) mass-energy densities and $w$ denotes the equation of state of the dark energy component. The dot denotes derivative with respect to conformal time. Let us note that for  $Q (t) > 0$, the energy transfer occurs from dark energy to dark matter, while  $Q(t) <0$  indicates the transfer of energy in the reverse direction, i.e.  from dark matter to dark energy.
From here one can see that the two dark components acquire a new behavior, that can be shown introducing the effective background equations of state of the dark matter and the dark energy components:
\begin{align}
    w_\mathrm{dm}^\mathrm{eff} &= -\frac{Q}{3H\rho_\mathrm{dm}}~, \label{eq:weffDM}\\
    w_\mathrm{de}^\mathrm{eff} &= w + \frac{Q}{3H\rho_\mathrm{de}}~.\label{eq:weffDE}
\end{align}

The most exploited parameterization of the interaction function is $Q = H \xi f(\rho_{\rm dm}, \rho_{\rm de})$~\cite{Gavela:2009cy}, where $\xi$ is the coupling parameter that characterizes the strength of the interaction function, $H$ is the Hubble parameter of the FLRW Universe and $f(\rho_{\rm dm}, \rho_{\rm de})$ is any continuous function of 
$\rho_{\rm dm}$, $\rho_{\rm de}$, i.e. of the energy densities of dark matter and dark energy. In this work we shall consider the following  well-known interaction function 
\begin{equation}
Q  =  H\,  \xi \, \rho_{\rm de}~, \label{eq:Q}
\end{equation}

\noindent 
where $\xi$ could be either time-dependent or time-independent. In the following, and for the sake of simplicity, we shall assume that such a coupling is time-independent. Now, in agreement with the sign convention of the interaction function, $\xi >0$ (equivalently, $Q > 0$) means an energy transfer from dark energy to dark matter and $\xi < 0$ means that the energy flow is from dark matter to dark energy. Notice that for some interaction models the energy density of dark matter and/or of dark energy could be negative~\cite{Pan_2020b}. 
In this regard we do not impose any further constraints on the fluids and let the data discriminate between the most observationally favoured scenarios. We believe this approach is appropriate because it avoids unwanted biases and also considers the rising interest in the community for the putative presence of a negative cosmological constant~\cite{Delubac_2015, Poulin_2018, Wang_2018, Visinelli_2019, Calder_n_2021}. We therefore extract the constraints on the cosmological  parameters, assuming the choice of the  interaction function given by Eq.~(\ref{eq:Q}). 

A  crucial parameter for deriving our constraints is the growth factor rate $f$, since RSD measurements~\cite{Hamilton_1998} constrain the combined quantity $f\sigma_8$, which is the product of the linear growth rate and the root mean square mass fluctuation amplitude for spheres of size $8\,h^{-1}\,\mathrm{Mpc}$. The mass variance of the matter clustering, for a generic physical scale $R$, is given by
\begin{equation}
\sigma^2_R=\frac{1}{2\pi^2}\int_0^{\infty}P(k,z)W_R^2(k)dk~,
\label{eq:sigma_R}
\end{equation}
\noindent where $P(k,z)$ is the matter power spectrum and $W_R(k)$ is the window function. The growth rate $f$ is defined as
\begin{equation}
f \equiv \frac{d\ln{\delta_\mathrm{dm}}}{d\ln{a}}=\frac{\delta'_\mathrm{dm}}{\delta_\mathrm{dm}}a~.
\label{eq:f_definition}
\end{equation}
Therefore, $f$ represents a measure of the evolution of the matter overdensity $\delta_\mathrm{dm}$ from the primordial density fluctuations to the large-scale structure observed today. The matter overdensity can be obtained by solving the so-called growth equation, which, within dark energy models characterized by a dark energy equation of state $w$, reads as  
\begin{equation}
\delta''_\mathrm{dm}=-(2-q)\frac{\delta'_\mathrm{dm}}{a}+\frac{3}{2}\Omega_\mathrm{dm}\frac{\delta_\mathrm{dm}}{a^2}  + \frac{3}{2}\Omega_\mathrm{b}\frac{\delta_\mathrm{b}}{a^2}~,
\label{eq:growth_equation_lcdm}
\end{equation}
\noindent where $q$ is the  deceleration parameter, defined as
\begin{equation}
q=-\frac{\dot{\mathcal{H}}}{\mathcal{H}^2}=\frac{1}{2}(1+3\,w\,\Omega_\mathrm{de})~.
\label{eq:deceleration_parameter}
\end{equation}
The growth rate $f$ can be parameterized as
\begin{equation}
f=[\Omega_\mathrm{dm}(a)]^\gamma=\left[\frac{\Omega_\mathrm{dm}a^{-3}}{\Omega_\mathrm{de}+\Omega_\mathrm{dm}a^{-3}}\right]^\gamma~,
\label{eq:f_parametrization}
\end{equation}
\noindent with $\Omega_\mathrm{dm}(a)$ the relative dark matter energy density and $\gamma$, called the growth index, is defined as a constant if we consider dark energy models within General Relativity. In particular, what one finds is that $\gamma\simeq3(w-1)/(6w-5)$~\cite{Linder_2007}. For the $\Lambda$CDM model, $w=-1$ and therefore $\gamma=6/11\simeq 0.55$~\cite{Linder_2005}.

\begin{figure}[h!]
\includegraphics[width=\textwidth]{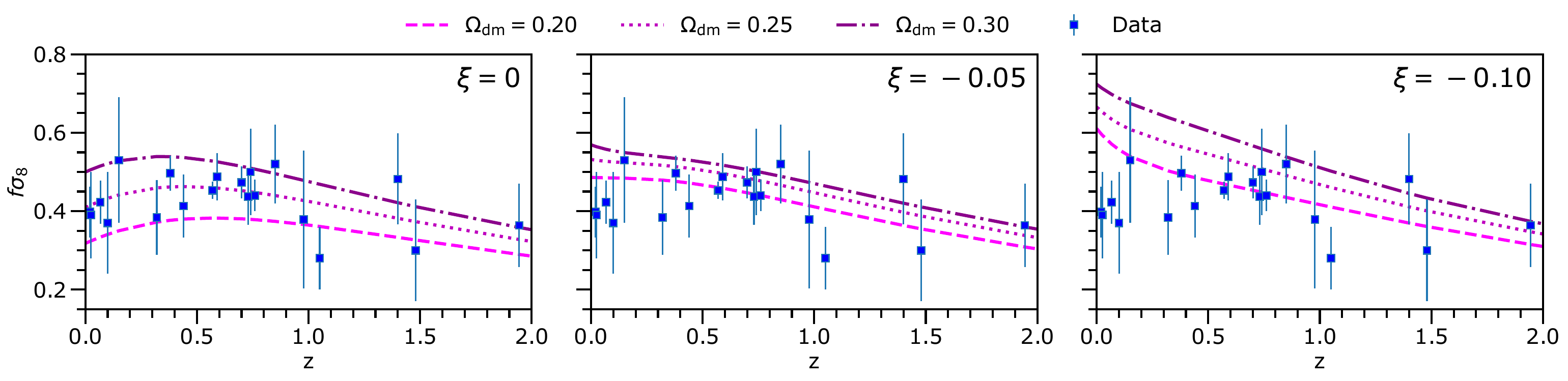}
\caption{\textbf{\emph{$\boldsymbol{f\sigma_8}$ as a function of the redshift}} for different values of $\Omega_\mathrm{dm}=0.20$, $0.25$ and $0.30$. The left, middle and right panel assume $\xi=0$ (i.e. $\Lambda$CDM), $-0.05$ and $-0.1$ respectively, see main text for details. The $f\sigma_8(z)$ data points are the ones included in Tab.~\ref{tab:TAB1_fs8_data_points} and correspond to the current total compilation of measurements of this quantity.}
\label{fig:FIG1_fs8_vs_z}
\end{figure}

For the interacting dark energy cosmologies considered here, the standard growth equation Eq.~(\ref{eq:growth_equation_lcdm}) is modified as
~\cite{Honorez_2010}
\begin{equation}
\delta''_\mathrm{dm}=-B\frac{\delta'_\mathrm{dm}}{a}+\frac{3}{2}A\Omega_\mathrm{dm}\frac{\delta_\mathrm{dm}}{a^2}  + \frac{3}{2}\Omega_b\frac{\delta_b}{a^2}~,
\label{eq:growth_equation_coupling}
\end{equation}
\noindent where
\begin{subequations}
\begin{align}
A&=1+\frac{2}{3}\frac{1}{\Omega_\mathrm{dm}}\frac{\rho_\mathrm{de}}{\rho_\mathrm{dm}}\left[-\xi(1-q-3w)+\xi^2\left(1+\frac{\rho_\mathrm{de}}{\rho_\mathrm{dm}}\right)\right] \label{eq:A_growth_equation}~;\\
B&=2-q+\xi\frac{\rho_\mathrm{de}}{\rho_\mathrm{dm}} \label{eq:B_growth_equation}~.
\end{align}
\end{subequations}
Figure~\ref{fig:FIG1_fs8_vs_z} shows the combined quantity $f\sigma_8$ versus the redshift for the standard  $\Lambda$CDM model and also for interacting dark sector cosmologies as the coupling strength increases. We also illustrate the dependence of $f\sigma_8$ on the dark matter energy density. The observational values of $f\sigma_8(z)$ are those reported in Tab.~\ref{tab:TAB1_fs8_data_points}. Notice that as the coupling strength increases, the discrepancy with the $\Lambda$CDM prediction also does, as well as the inconsistency with the majority of the observational data points, especially for large values of the dark matter energy density. That is, for a fixed value of the dark matter energy density today, the departure with the standard growth of structure picture raises as the value of the coupling gets more negative.

\section{Methodology and data sets}
\label{sec:method}
\subsection{Cosmological measurements}

In this section we present the datasets and likelihoods used to derive the constraints on our interacting dark matter-dark energy model.

\textbf{\emph{CMB measurements}$\,-\,$} The Planck mission~\cite{planck18VI, planck18I, planck18V} has achieved exceptionally precise measurements of the power spectra of CMB anisotropies. The CMB power spectra contains a vast amount of information. We use as our baseline dataset the temperature (TT) and polarization (EE) auto-spectra, plus their cross-spectra (TE), as incorporated in the \texttt{Commander} (for multipoles $\ell<30$) and \texttt{plik} (for multipoles $\ell>30$) likelihoods from the PR3 release~\cite{planck18V}. In addition to the primary temperature and polarization anisotropy power spectra, we also have information on the power spectrum of the gravitational lensing potential~\cite{planck18VIII}. All of the likelihoods described above are already included in \texttt{Cobaya}~\cite{2019cobaya, Torrado_2021}. In the following, we shall denote with \textbf{\textit{Planck}} the results obtained using Planck temperature, polarization and lensing measurements.

\textbf{\emph{RSD measurements}$\,-\,$} 
As RSD reference measurements, we consider the data points in Tab.~\ref{tab:TAB1_fs8_data_points} taken from Ref.~\cite{Avila_2022}. If two or more sets of data concern the same cosmological tracer, the considered $[f\sigma_8](z)$ data are obtained from uncorrelated redshift bins. On the other hand, when different cosmological tracers were analysed, the used data is taken from possibly correlated redshift. When the same galaxy survey  performed two or more measurements corresponding to different data releases, we consider only the latest measurement of $f\sigma_8$. We built our own likelihood since the main observations exploited here are not included in \texttt{Cobaya}.  In the following, we shall denote this dataset as \textbf{\textit{RSD}}.
\begin{table}[h!]\scriptsize
\setlength{\tabcolsep}{10pt}
\renewcommand{\arraystretch}{1.5}
\begin{center}
\begin{tabular}{|c||c|c|c|c|}
\hline
\textbf{Survey} & $\boldsymbol{z}$ & $\boldsymbol{f\sigma_8}$ & \textbf{Reference} & \textbf{Cosmological tracer} \\ \hline
SnIa + IRAS & 0.02 & 0.398 $\pm$ 0.065 &  \cite{Turnbull_2011} & SNIa + galaxies \\
6dFGS & 0.025 & 0.39 $\pm$ 0.11 &  \cite{Achitouv_2017} & void \\
6dFGS & 0.067 & 0.423 $\pm$ 0.055 &  \cite{Beutler_2012} & galaxies \\
SDSS-veloc & 0.10 & 0.37 $\pm$ 0.13 &  \cite{Feix_2015} & DR7 galaxies \\
SDSS-IV & 0.15 & 0.53 $\pm$ 0.16 &  \cite{Alam_2017} & eBOSS DR16 MGS \\
BOSS-LOWZ & 0.32 & 0.384 $\pm$ 0.095 &  \cite{S_nchez_2014} & DR10, DR11 \\
SDSS-IV & 0.38 & 0.497 $\pm$ 0.045 &  \cite{Alam_2017} & eBOSS DR16 galaxies \\
WiggleZ & 0.44 & 0.413 $\pm$ 0.080 &  \cite{Blake_2012} & LRG \& bright emission-line galaxies \\
CMASS-BOSS & 0.57 & 0.453 $\pm$ 0.022 &  \cite{Nadathur_2019} & DR12 voids + galaxies \\
SDSS-CMASS & 0.59 & 0.488 $\pm$ 0.060 &  \cite{Chuang_2016} & DR12 \\
SDSS-IV & 0.70 & 0.473 $\pm$ 0.041 &  \cite{Alam_2017} & eBOSS DR16 LRG \\
WiggleZ & 0.73 & 0.437 $\pm$ 0.072 &  \cite{Blake_2012} & bright emission-line galaxies \\
SDSS-IV & 0.74 & 0.50 $\pm$ 0.11 &  \cite{Aubert_2022} & eBOSS DR16 voids \\
VIPERS v7 & 0.76 & 0.440 $\pm$ 0.040 &  \cite{wilson2016geometric} & galaxies \\
SDSS-IV & 0.85 & 0.52 $\pm$ 0.10 &  \cite{Aubert_2022} & eBOSS DR16 voids \\
SDSS-IV & 0.978 & 0.379 $\pm$ 0.176 &  \cite{Zhao_2018} & eBOSS DR14 quasars \\
VIPERS v7 & 1.05 & 0.280 $\pm$ 0.080 &  \cite{wilson2016geometric} & galaxies \\
FastSound & 1.40 & 0.482 $\pm$ 0.116 &  \cite{Okumura_2016} & ELG \\
SDSS-IV & 1.48 & 0.30 $\pm$ 0.13 &  \cite{Aubert_2022} & eBOSS DR16 voids \\
SDSS-IV & 1.944 & 0.364 $\pm$ 0.106 &  \cite{Zhao_2018} & eBOSS DR14 quasars \\\hline
\end{tabular}
\end{center}
\caption{\textbf{\emph{Compilation of $\boldsymbol{[f\sigma_8](z)}$ measurements}} obtained from Ref.~\cite{Avila_2022}.}
\label{tab:TAB1_fs8_data_points}
\end{table}

\textbf{\emph{Supernovae measurements}$\,-\,$}
Type Ia Supernovae (SN Ia) serve as standardizable candles that can be used to measure the expansion of the Universe. Within the $\Lambda$CDM model, SN Ia have lower statistical power with respect to modern BAO measurements, but are still useful to reduce the number of degeneracies inside the cosmological parameter set. 
The SN Ia dataset considered in this work is the Pantheon+ compilation~\cite{Scolnic_2022}. It contains 1550 spectroscopically confirmed SN Ia in the redshift range $0.001<z<2.26$. 
We use the public likelihood of Ref.~\cite{Brout_2022}, included in \texttt{Cobaya}. In the following, we shall denote this dataset as \textbf{\textit{PantheonPlus}}.

\textbf{\emph{DESI measurements}$\,-\,$}
DESI spectroscopic targets are selected from photometric catalogs of the 9th public data release of the DESI Legacy Imaging Surveys~\cite{desiVI}. The five tracer samples, covering a total redshift range from $z=0.1$ to $z=4.2$, are: 
Bright Galaxy Samples (BGS)~\cite{Hahn_2023} in the range $0.1<z<0.4$, Luminous Red Galaxy Sample (LRG)~\cite{Zhou_2023} in the range $0.4<z<0.6$ and $0.6<z<0.8$, Emission Line Galaxy Sample (ELG)~\cite{Raichoor_2023} in the range $1.1<z<1.6$, combined LRG and ELG Sample (LRG+ELG) in the range $0.8<z<1.1$~\cite{desiIII}, Quasar Sample (QSO)~\cite{Chaussidon_2023} in the range $0.8<z<2.1$ and Lyman-$\alpha$ Forest Sample (Ly$\alpha$)~\cite{desiIV} in the range $1.77<z<4.16$. In the following, we shall denote these BAO datasets as \textbf{\textit{DESI}}.

\subsection{Cosmological inference}

The Boltzmann solver used to interface with the cosmological inference code (\texttt{Cobaya}~\cite{2019cobaya, Torrado_2021}) is a modified version of the Cosmic Linear Anisotropy Solving System code (\texttt{CLASS})~\cite{lesgourgues2011, Diego_Blas_2011}, in order to include the effect of the coupling between the dark sectors. 
The Bayesian inference is performed using the Metropolis-Hastings MCMC sampler~\cite{Lewis_2002, Lewis_2013} implemented in \texttt{Cobaya}~\cite{2019cobaya, Torrado_2021}. In order to test the convergence of the chains obtained using this approach, we utilize the Gelman-Rubin criterion~\cite{gelmanrubin}, and we establish a threshold for chain convergence of $R-1 \leq 0.01$. We sample the set of parameters of our extension of the $\Lambda\mathrm{CDM}$ model including the dark coupling (referred from now on as $\Lambda\mathrm{CDM}+\xi$), i.e. \{$w_\mathrm{b}, w_\mathrm{dm}, 100\theta_\mathrm{s}, \ln(10^{10}A_\mathrm{s}), n_\mathrm{s}, \tau, \xi$\}, where $w_{\rm{b(dm)}}=\Omega_\mathrm{b(dm)}h^2$ is the baryon (cold dark matter) energy density, $\theta_\mathrm{s}$ is the angular size of the horizon at the last scattering surface, $\tau$ is the optical depth, $\log(10^{10}A_\mathrm{s})$ is the amplitude of primordial scalar perturbation and $n_\mathrm{s}$ is the scalar spectral index. Table~\ref{tab:TAB2_priors} presents the priors used for all the seven sampled parameters in the MCMC analyses in the $\Lambda\mathrm{CDM}+\xi$ fiducial cosmology explored here. All of the priors are uniform distributions in the given ranges. We considered the equation of state parameter for dark energy, $w$, to be $w=-1+\epsilon$, with $\epsilon = 0.01$ in order to regularise early-time super-horizon instabilities in the dynamics of cosmological perturbations~\cite{Gavela:2009cy, V_liviita_2008}.

\begin{table}[h!]
\centering
\setlength{\tabcolsep}{10pt}
\renewcommand{\arraystretch}{1.5}
\resizebox{\textwidth}{!}{
\begin{tabular} {|c||c c c c c c c|}
\hline
Parameter & {\boldmath$\log(10^{10} A_\mathrm{s})$} & {\boldmath$n_\mathrm{s}   $} & {\boldmath$100\theta_\mathrm{s}$} & {\boldmath$\Omega_\mathrm{b} h^2$} & {\boldmath$\Omega_\mathrm{dm} h^2$} & {\boldmath$\tau_\mathrm{reio}$} & {\boldmath$\xi$} \\
\hline 
Prior & $\mathcal{U}[1.61, 3.91]$ & $\mathcal{U}[0.8, 1.2]$ & $\mathcal{U}[0.5, 10]$ & $\mathcal{U}[0.005, 0.1]$ & $\mathcal{U}[0.001, 0.99]$ & $\mathcal{U}[0.01, 0.8]$ & $\mathcal{U}[-1.0, 0.0]$ \\
\hline 
\end{tabular}}
\caption{\textbf{\emph{Cosmological parameters and prior input}} for each parameter sampled in our bayesian parameter inference. The $\mathcal{U}[a,b]$ nomenclature denotes a uniform distribution with lower limit $a$ and upper limit $b$.} 
\label{tab:TAB2_priors}
\end{table}

\section{Constraints on the dark sector coupling}
\label{sec:results}
In the following we shall present the constraints coming from the combination of different cosmological measurements. The combinations of the previously described datasets considered are:

\begin{itemize}
    \item \textbf{\textit{Planck}};
    \item \textbf{\textit{Planck}} + \textbf{\textit{RSD}};
    \item \textbf{\textit{Planck}} + \textbf{\textit{DESI}};
    \item \textbf{\textit{Planck}} + \textbf{\textit{DESI}} + \textbf{\textit{RSD}};
    \item \textbf{\textit{Planck}} + \textbf{\textit{DESI}} + \textbf{\textit{PantheonPlus}} + \textbf{\textit{RSD}}.
\end{itemize}

\begin{table}[h!]
\centering
\setlength{\tabcolsep}{10pt}
\renewcommand{\arraystretch}{1.5}
\resizebox{\textwidth}{!}{
\begin{tabular} {|c||c c c c c|}
\hline
\multirow{2}{*}{Parameter} & \multirow{2}{*}{\textbf{\textit{Planck}}} &\multirow{2}{*}{\textbf{\textit{Planck $+$ RSD}}} & \multirow{2}{*}{\textbf{\textit{Planck $+$ DESI}}} & \textbf{\textit{Planck $+$ DESI}} & \textbf{\textit{Planck $+$ DESI}} \\
 &  &  &  & \textbf{\textit{$+$ RSD}} & \textbf{\textit{$+$ PantheonPlus $+$ RSD}} \\
\hline
{\boldmath$\log(10^{10} A_\mathrm{s})$} & $3.044\pm 0.014$ & $3.043^{+0.015}_{-0.013}$ & $3.048\pm 0.014$ & $3.050^{+0.013}_{-0.015}$ & $3.047\pm 0.014$\\
{\boldmath$n_\mathrm{s}$} & $0.9657\pm 0.0042$ & $0.9654\pm 0.0041$ & $0.9674\pm 0.0037$ & $0.9692\pm 0.0036$ & $0.9684\pm 0.0036$\\
{\boldmath$100\theta_\mathrm{s}$} & $1.04190\pm 0.00030$ & $1.04188\pm 0.00029$ & $1.04198\pm 0.00028$ & $1.04204\pm 0.00028$ & $1.04200\pm 0.00028$\\
{\boldmath$\Omega_\mathrm{b} h^2$} & $0.02238\pm 0.00015$ & $0.02239\pm 0.00015$ & $0.02244\pm 0.00014$ & $0.02251\pm 0.00013$ & $0.02247\pm 0.00013$\\
{\boldmath$\Omega_\mathrm{dm} h^2$} & $0.066^{+0.052}_{-0.023}$ & $0.1184^{+0.0017}_{-0.0014}$ & $0.086^{+0.029}_{-0.012}$ & $0.1167^{+0.0016}_{-0.0011}$ & $0.1172^{+0.0014}_{-0.0010}$\\
{\boldmath$\tau_\mathrm{reio}$} & $0.0541\pm 0.0075$ & $0.0539\pm 0.0072$ & $0.0569\pm 0.0073$ & $0.0589^{+0.0065}_{-0.0079}$ & $0.0572\pm 0.0072$\\
{\boldmath$\xi$} & $> -0.565$ & $> -0.0134$ & $-0.268^{+0.26}_{-0.083}$ & $> -0.0155$ & $> -0.0136$\\
\hline
\end{tabular}}
\caption{\textbf{\emph{68\%~CL intervals for the IDE model}} obtained using our modified version of \texttt{CLASS} for all the data combinations we have considered. We add the coupling $\xi$ as an additional parameter. All parameters are sampled in the MCMC analysis with flat priors.}
\label{tab:TAB3_results}
\end{table}

\begin{figure}[h!]
    \centering
    \includegraphics[width=0.98\textwidth]{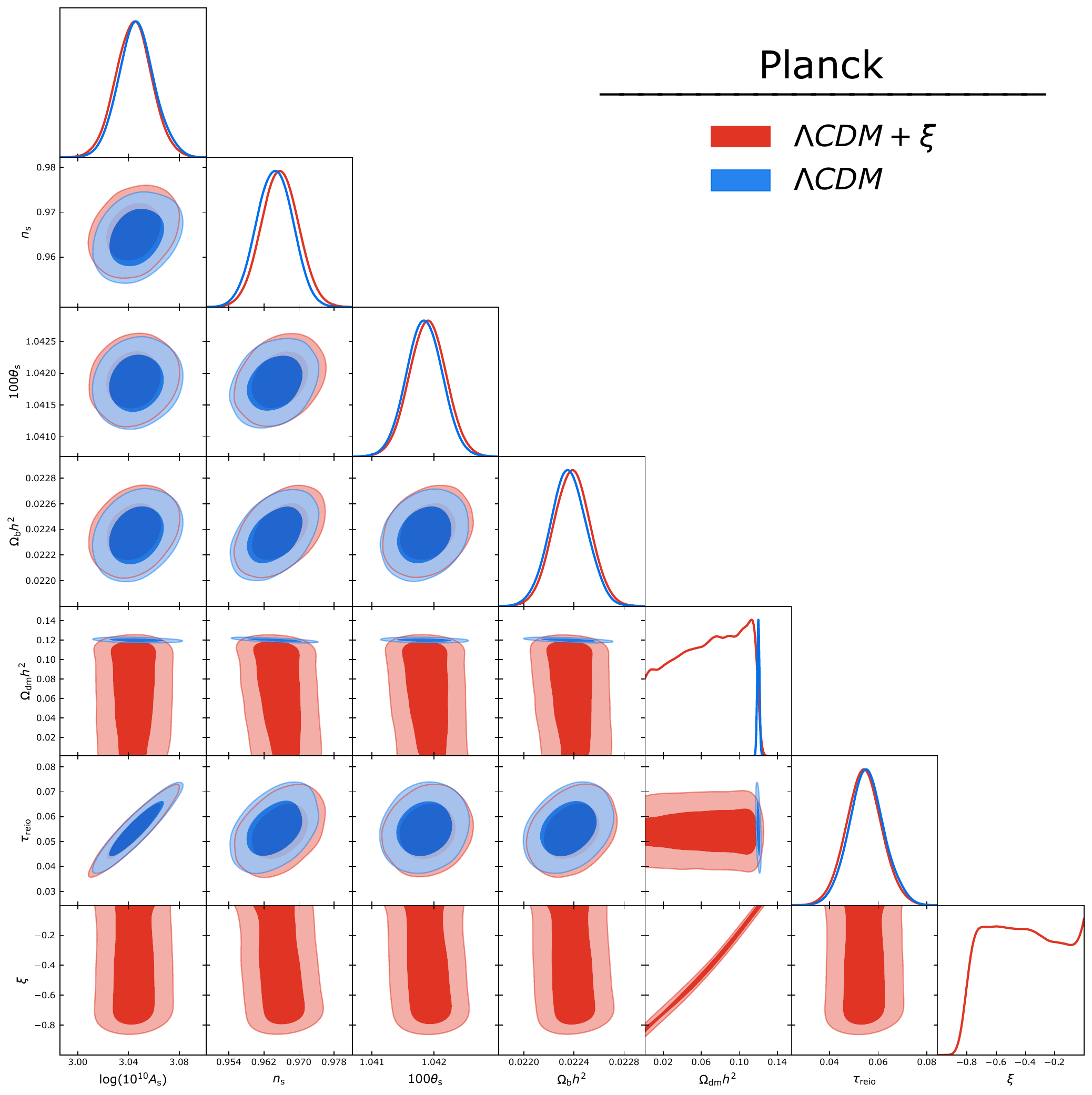}
    \caption{\textbf{\emph{Planck 68\% and 95\%~CL constraints on the parameters of our $\boldsymbol{\Lambda\mathrm{CDM}+\xi}$ model.}} We use the combination of TT, EE, TE and lensing spectra of Planck. We show how the results for our model (red regions) compare to those obtained in the $\Lambda\mathrm{CDM}$ case (blue regions).}
    \label{fig:FIG2_planck}
\end{figure}

We start by describing the constraints from  Planck CMB data alone. The effect of the introduced coupling between dark matter and dark energy, as expected, has a significant influence mostly on the $\Omega_\mathrm{dm}h^2$ parameter. Indeed, as one can see from Fig.~\ref{fig:FIG2_planck}, there is a strong degeneracy between $\Omega_\mathrm{dm}h^2$ and $\xi$, with lower values of $\Omega_\mathrm{dm}h^2$ being permissible (with respect to the $\Lambda$CDM results) for more negative values of the coupling $\xi$. As a consequence, from Tab.~\ref{tab:TAB3_results}, we can notice that the best-fit value for $\Omega_\mathrm{dm}h^2$ is significantly lower with respect to the $\Lambda$CDM one $\Omega_\mathrm{dm}h^2|_\mathrm{\Lambda CDM}=0.1200$. All the other parameters, instead, appear to be less sensitive to the introduced coupling between dark energy and dark matter, as the obtained results are consistent with those obtained in the $\Lambda$CDM case (see Fig.~\ref{fig:FIG2_planck}). These results are consistent with the fact that Planck data alone are not able to break the strong degeneracy between the dark matter energy density and the coupling. 

\begin{figure}[h!]
    \centering
    \includegraphics[width=0.98\textwidth]{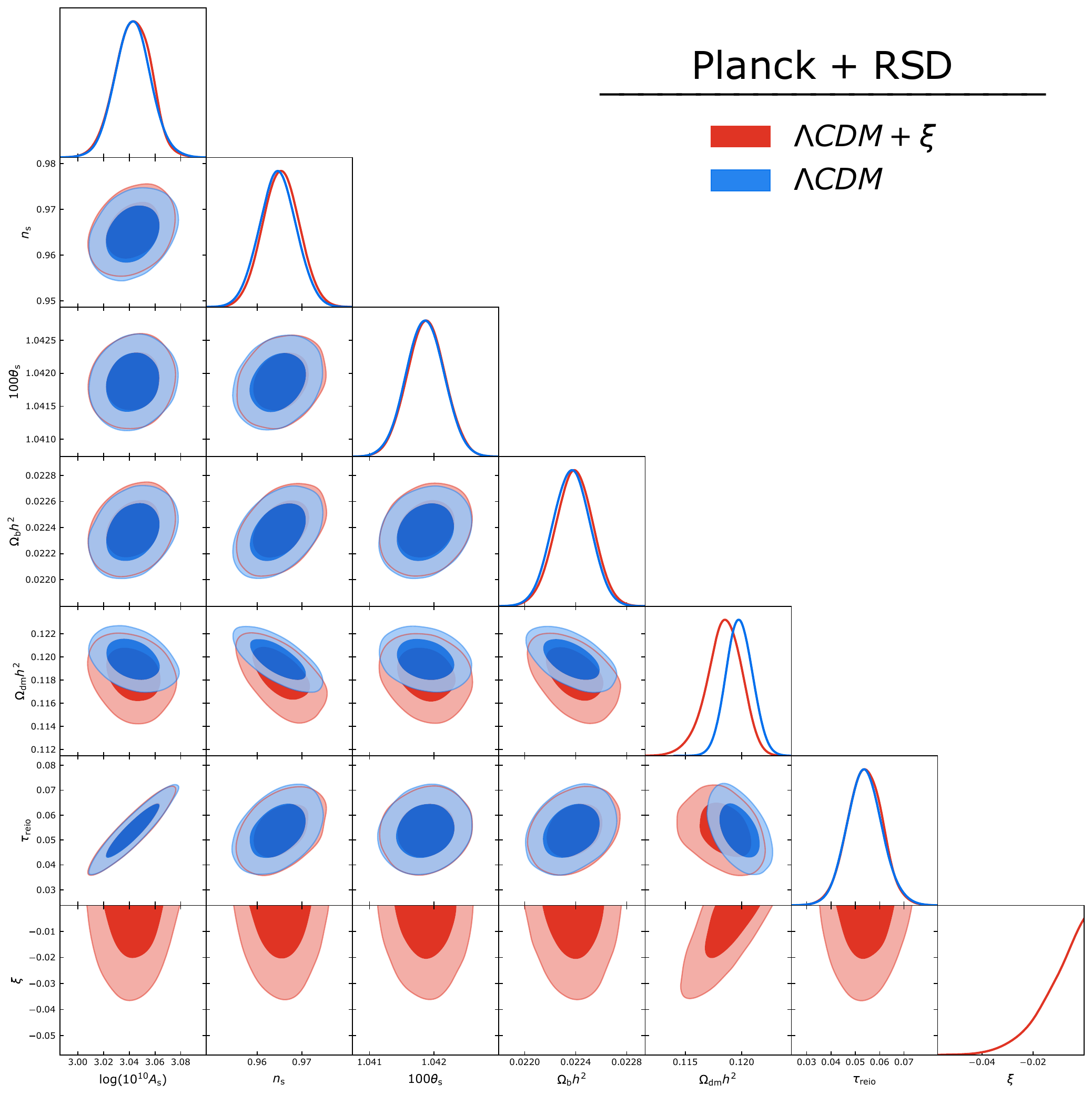}
    \caption{\textbf{Planck+RSD \emph{68\% and 95\%~CL constraints on the parameters of our $\boldsymbol{\Lambda\mathrm{CDM}+\xi}$ model.}} We use the combination of TT, EE, TE and lensing spectra of Planck, together with redshift space distortions measurements. We show how the results of our model (red regions) compare to those obtained in the $\Lambda$CDM case (blue regions).}
    \label{fig:FIG3_planck_rsd}
\end{figure}

Once we consider RSD observations, despite the fact that also in this case the only parameter affected by the presence of the coupling is $\Omega_\mathrm{dm}h^2$, differently from the results obtained considering Planck alone, we notice that the RSD measurements are indeed very constraining. Figure~\ref{fig:FIG3_planck_rsd} makes clear this point, as the degeneracy between the dark matter energy density and the dark coupling is notably reduced with respect to the case of Planck data alone. Indeed, we notice that the best-fit value of $\Omega_\mathrm{dm}h^2$ quoted in Tab.~\ref{tab:TAB3_results} matches the result obtained in the standard cosmological model using Planck data only. These results are consistent with the fact that RSD measurements constrain the late time physics of the Universe, when the dominant effects induced by the coupling $\xi$ should become more important. RSD measurements help to break the degeneracy between $\xi$ and $\Omega_\mathrm{dm}h^2$, placing a tighter constraint on the coupling $\xi$. When Planck data are combined with RSD measurements, the bounds mainly improve on the parameters related to the growth rate of structure. 

\begin{figure}[h!]
    \centering
    \includegraphics[width=0.98\textwidth]{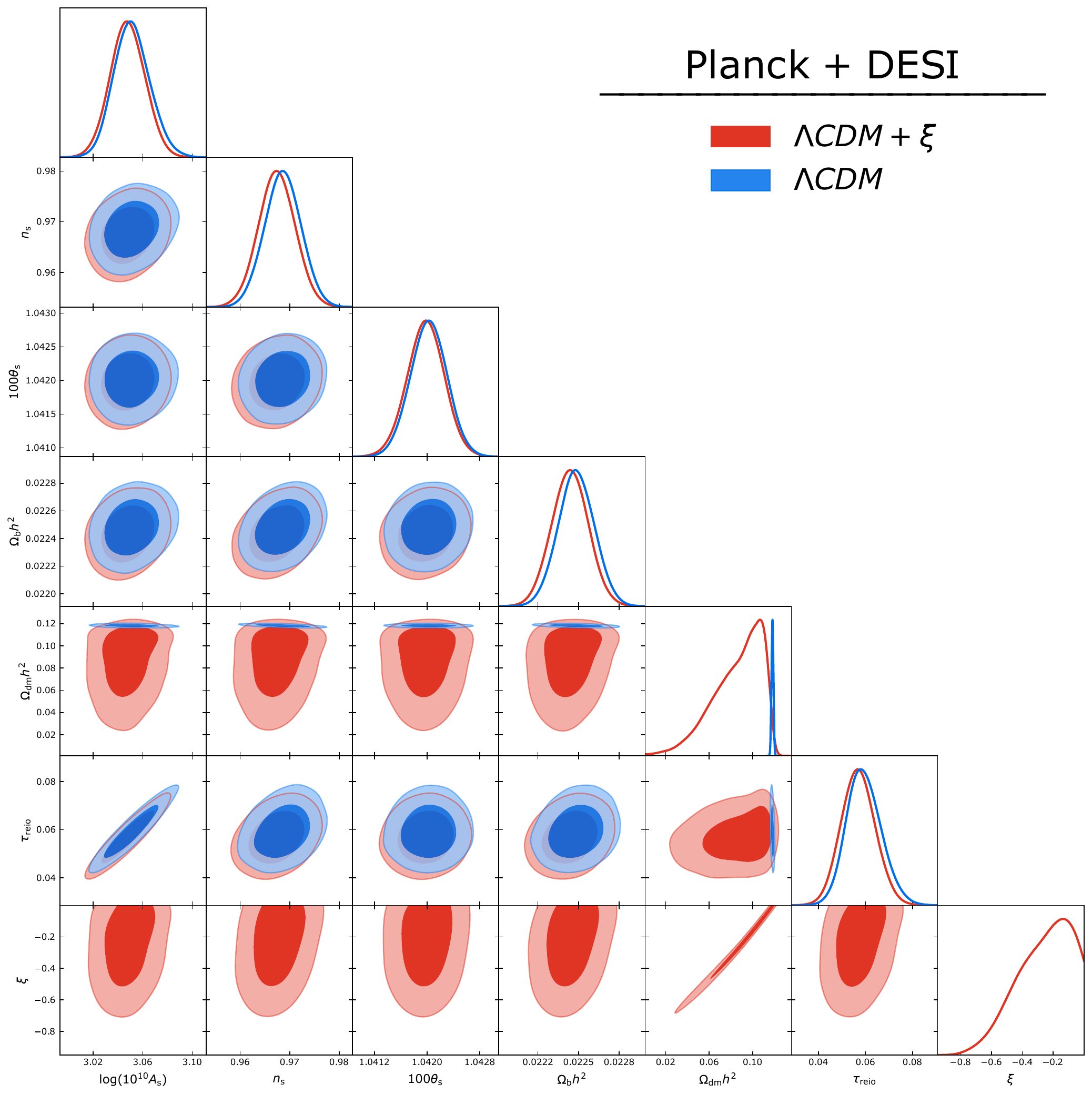}
    \caption{ \textbf{\emph{Planck+DESI 68\% and 95\%~CL constraints on the parameters of our $\boldsymbol{\Lambda\mathrm{CDM}+\xi}$ model.}} We use the combination of TT, EE, TE and lensing spectra of Planck, together with BAO measurements from DESI.  We show how the results of our model (red regions) compare with those obtained in the $\Lambda$CDM case (blue regions).}
    \label{fig:FIG4_planck_desi}
\end{figure}

The following data combination we describe here is Planck plus DESI BAO observations. As can be noticed from Fig.~\ref{fig:FIG4_planck_desi}, the degeneracy between $\Omega_\mathrm{dm} h^2$ and $\xi$ is the main effect of the coupling, as all other parameters perfectly reproduce the results obtained in the $\Lambda$CDM case. Interestingly, despite the fact that both RSD measurements and the DESI BAO measurements provide information on late time physics, RSD measurements have a much more constraining power than DESI BAO observations, as the strong degeneracy between $\Omega_\mathrm{dm} h^2$ and $\xi$ is more effectively alleviated in the case of the combination of CMB data with RSD.  We also notice that the combination of Planck and DESI seems to prefer a more negative value of the coupling with respect to the case of Planck and RSD. This is due to the slightly larger value of the Hubble constant preferred by DESI observations. Notice that the best-fit for $\xi$ differs from zero, we refer the reader to the recent work of Ref.~\cite{giarè2024} for further discussion. Nevertheless, in order to assess this neat result one should wait for future DESI data releases. Based on the results of DESI observations reported in~\cite{desiVI}, it seems that a non canonical value for the  equation of state of the dark energy component is preferred. Interacting dark energy cosmologies, indeed, introduce an effective equation of state for dark energy (see eq.~(\ref{eq:weffDE})), that differs from the standard cosmological constant value, and such a dark energy equation of state may be preferred by DESI measurements, showing therefore a mild preference for a non-zero coupling. 

\begin{figure}[h!]
    \centering
    \includegraphics[width=0.98\textwidth]{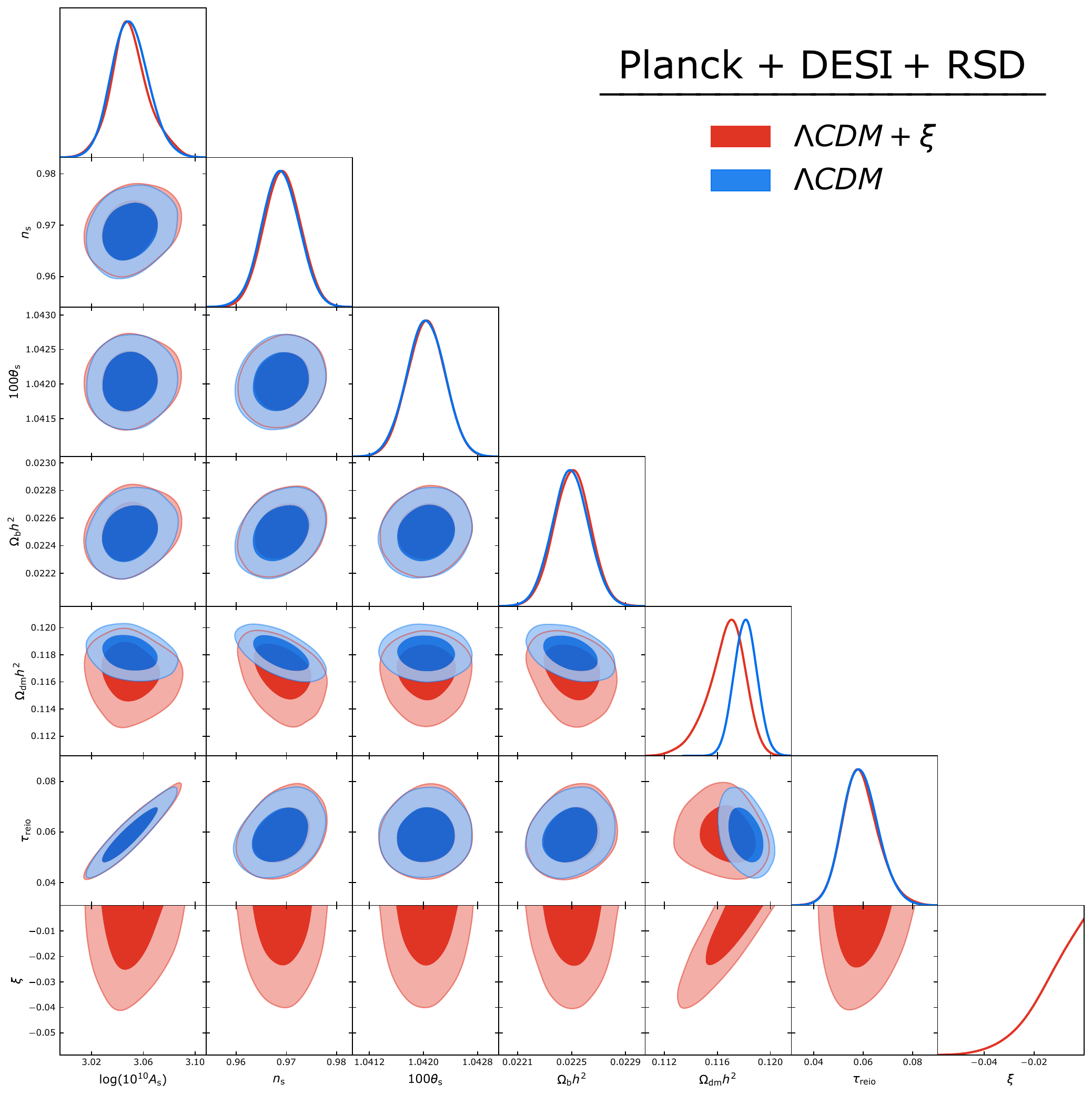}
    \caption{\textbf{\emph{Planck+DESI+RSD 68\% and 95\%~CL constraints on the parameters of our $\boldsymbol{\Lambda\mathrm{CDM}+\xi}$ model.}} We use the combination of TT, EE, TE and lensing spectra of Planck, together with BAO measurements from DESI and redshift space distortions measurements.  We show how the results of our model (red regions) compare with those obtained in the $\Lambda$CDM case (blue regions).}
    \label{fig:FIG5_planck_desi_rsd}
\end{figure}

As previously stated, RSD measurements appear to have a much stronger effect on the constraints on $\xi$, thanks to the fact that they directly measure the impact of the interaction rate on the growth of structure. 
Figure~\ref{fig:FIG5_planck_desi_rsd} illustrates the fact that the bounds on the coupling when adding RSD to Planck+DESI changes are led by the strong constraining power of the RSD dataset. The degeneracy of $\Omega_\mathrm{dm}h^2$ with the coupling $\xi$ is greatly reduced due to the tighter lower bound on $\xi$, making the value of the $\Omega_\mathrm{dm}h^2$ parameter very similar to its $\Lambda$CDM result (see Tab.~\ref{tab:TAB3_results}).

\begin{figure}[h!]
    \centering
    \includegraphics[width=\textwidth]{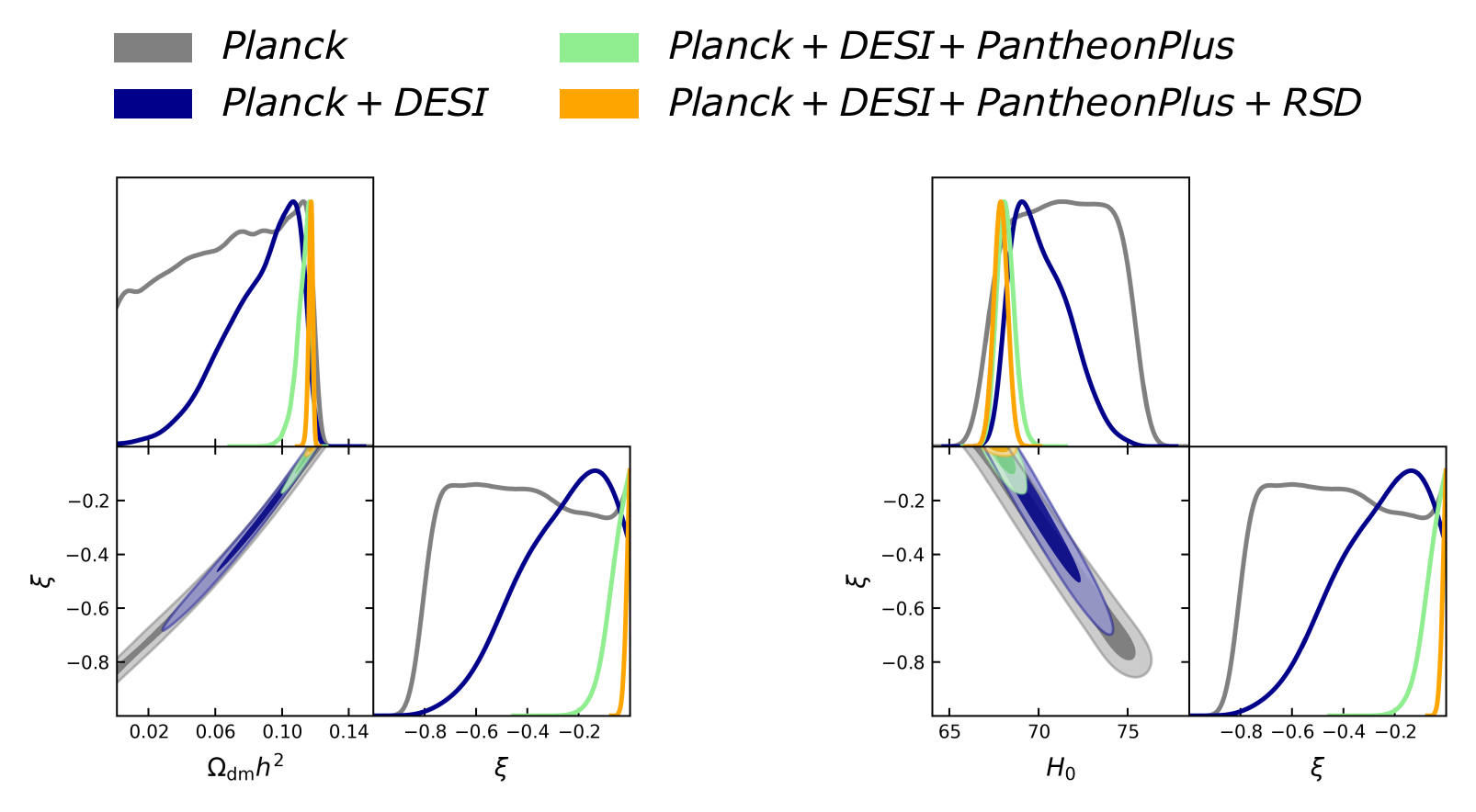}
    \caption{\textbf{\emph{Degeneracy between the $\boldsymbol{\{\xi,\,\Omega_{\rm dm}h^2\}}$ and $\boldsymbol{\{\xi,\,H_0\}}$ parameters.}} In the figure we show the one-dimensional posterior distributions and 68\% and 95\%~CL constraints on $\xi$ and $H_0$ (right panel) and on $\xi$ and $\Omega_{\rm dm}h^2$ (left panel) for our $\Lambda\mathrm{CDM}+\xi$ model using different combinations of measurements.  All the results are obtained in the $\Lambda\mathrm{CDM} + \xi$ scenario. From both panels we can see how RSD measurements corner the value of the coupling to a very small value, close to the standard $\Lambda\mathrm{CDM}$ model ($\xi=0$).}
    \label{fig:FIG6_xi_vs_odm_H0}
\end{figure}

Figure~\ref{fig:FIG6_xi_vs_odm_H0} illustrates the strong degeneracy between the Hubble constant $H_0$ and the coupling $\xi$. Indeed, as previously mentioned, IDE models could be able to alleviate the so-called $H_0$ tension, see Refs.~\cite{Di_Valentino_2017, Di_Valentino_2020, Di_Valentino_2020a, Yang_2021a, Yang_2020d, Pan_2019, Yang_2019b, Yang_2018c, Yang_2018d}. We notice that Planck alone allows higher values of $H_0$, consistent with the measurements of SH0ES. This is related to the very strong degeneracy that appears between $\Omega_\mathrm{dm}h^2$ and the coupling $\xi$ (see Fig.~\ref{fig:FIG2_planck}), that is translated into a degeneracy between $H_0$ and $\xi$. Also including DESI measurements, we notice that there is still a huge degeneracy, which reduces the $H_0$ tension as for more negative couplings, higher $H_0$ values are allowed. As soon as we consider also Supernovae and RSD measurements, the constraints become tighter. In particular, considering the right panel of Fig.~\ref{fig:FIG6_xi_vs_odm_H0}, we notice that the RSD measurements give a very tight constraint for the coupling, which translates into a value for the Hubble parameter today in agreement with the one we have from Planck. Therefore, focusing on Planck and Planck+DESI measurements, interacting dark energy models could fully solve the $H_0$ tension. We remind the reader that actually the addition of DESI data seems to prefer a non zero coupling, even though still compatible with a zero value for $\xi$ at the 95\% CL. When one includes also SN Ia measurements, we obtain lower values of $H_0$ with respect to the previous dataset combinations considered, but higher than those obtained with the CMB measurements of Planck in the $\Lambda$CDM scenario. This alleviates the $H_0$ tension as well, even if not significantly. It is only when we introduce the RSD measurements that the constraints become extremely tight, favoring values of $H_0$ which are in agreement with the Planck $\Lambda$CDM results, not allowing to alleviate the tension. This is a neat important result, since the RSD measurements are the most sensitive observable to the effect of the coupling. The left panel of  Fig.~\ref{fig:FIG6_xi_vs_odm_H0} compares how the expected degeneracy between $\Omega_\mathrm{dm}h^2$ and $\xi$ gets constrained by the different measurements. Here, the behaviour is the same as before. Since $\xi$ and $\Omega_\mathrm{dm}h^2$ are strongly degenerate, a tighter constraint on the coupling $\xi$ will lead to a tighter constraint in $\Omega_\mathrm{dm}h^2$. We notice that with Planck data alone the degeneracy is very strong, which is consistent with the fact that CMB measurements are less sensitive to late time physics. Including DESI measurements, we notice that the degeneracy gets mildly reduced. As soon as we consider also Supernovae and RSD measurements, the constraints becomes tighter. In particular, we notice that the RSD measurements give a very tight constraint for the coupling, which translates into a smaller degeneracy in the ($\Omega_\mathrm{dm}h^2$, $\xi$) plane.

\section{Conclusions}
\label{sec:conclusions}
 We have investigated the role of the growth of structure in constraining non-standard dark energy cosmologies, focusing on an interacting dark energy (IDE) scenario, in which the interaction is proportional to the dark energy energy density. We place constraints on the phenomenological parameters of these alternative cosmological models by analyzing various cosmological datasets, in different combinations, to obtain a complementary analysis. The cosmological analyses yield constraints consistent with a zero coupling scenario ($\xi \sim 0$), considering all the datasets taken into account in this work. These findings indicate that current cosmological observations do not yet provide significant evidence for non-gravitational interactions between dark matter and dark energy. As one would expect, Planck observations are not able to set very strong constraints on the coupling, as  they are less sensible to the late time physics of the Universe and therefore there is a strong degeneracy between the coupling and the dark matter energy density when considering CMB data only. When adding all the other datasets, the constraints become tighter. This is because Supernovae, BAO and RSD measurements are more sensitive to the time in which dark energy and the flow between the two dark sectors  becomes important. Nevertheless, some of these late time observables are more sensitive than the others, being RSD the ones providing the tighter constraints. The most interesting results are those obtained introducing the DESI BAO measurements, which lead to a non-zero best-fit value for the coupling, even if still consistent with a vanishing value for $\xi$ at 95\% CL. In order to fully assess this result, one should wait for future releases from the DESI collaboration. We find, as we would expect, that the largest degeneracy of the coupling $\xi$ appears with the density parameter of cold dark matter, $\Omega_\mathrm{dm}h^2$. The degeneracy between these two parameters is strong  when considering only CMB measurements. Nevertheless, when we introduce other measurements, which are more sensitive to late time physics (when $\xi$ becomes more important), the degeneracy is reduced. Our analyses place particular emphasis on RSD measurements, which have not been extensively highlighted in the recent literature. RSD data are crucial: indeed, we found that it is only when we add them that the constraints become much tighter, helping enormously in breaking the degeneracies. Our findings have implications for the $H_0$ tension, which suggests discrepancies in the Hubble constant values derived from different observational methods. For the same reason above, while some measurements present a huge degeneracy between $H_0$ and $\xi$, alleviating the tension, when we include the RSD data the degeneracy almost disappears and the constraints on the Hubble constant are restored to the Planck $\Lambda$CDM values. Even though the cosmological analyses including all the data sets considered here point towards a vanishing coupling scenario, indicating that the interacting dark energy cosmologies could not have the final answer to the $H_0$ tension, one must keep in mind that we only focused in this work on a particular case, among all the possible ones. In summary, while current observations favor a zero dark energy-dark matter coupling scenario, the use of cosmological observations, including RSD measurements, lay the groundwork for future investigations into the dynamics of the Universe. Expanding the range of models and incorporating future datasets will be essential in uncovering the true nature of dark energy, dark matter, and their potential interactions.
\begin{acknowledgments}

The authors would like to thank William Giarè for enlightening discussions. This work has been supported by the Spanish MCIN/AEI/10.13039/501100011033 grants PID2020-113644GB-I00 (RH and OM) and by the European Union’s Horizon 2020 research and innovation programme under the Marie Skłodowska-Curie grants H2020-MSCA-ITN-2019/860881-HIDDeN and HORIZON-MSCA-2021-SE-01/101086085-ASYMMETRY. RH is supported by the Spanish grant FPU19/03348 of MU.
The authors also acknowledge support from the Generalitat Valenciana grants PROMETEO/2019/083 and CIPROM/2022/69 (RH and OM). RH and OM acknowledge the Galileo Galilei Institute for Theoretical Physics (GGI) for their hospitality during the completion of this work. OM acknowledges the financial support from the MCIU with funding from the European Union NextGenerationEU (PRTR-C17.I01) and Generalitat Valenciana (ASFAE/2022/020).
\end{acknowledgments}

\bibliography{bib}
\end{document}